\documentclass[a4paper, amsfonts, amssymb, amsmath, reprint, showkeys, nofootinbib, twoside,superscriptaddress]{revtex4-2}
\usepackage[english]{babel}
\usepackage[utf8]{inputenc}
\usepackage[colorinlistoftodos, color=green!40, prependcaption]{todonotes}
\usepackage{siunitx}
\usepackage{svg}
\usepackage{graphicx}
\usepackage[position=top]{subfig}
\usepackage[T1]{fontenc}
\newcommand\thefont{\expandafter\string\the\font}
\usepackage{amsmath}
\usepackage{mathtools}
\usepackage{mathrsfs}
\usepackage{mathptmx}
\usepackage{float}

\renewcommand{\dot}[1] {\overset{\,_{\mbox{\large \bf .}}}{#1}}

\begin{document}
%\title{Resonance frequency tracking with frequency counters for nanomechanical sensors}
\title{Self-Sustaining Oscillator with Frequency Counter for Resonance Frequency Tracking \\in Micro- and Nanomechanical Sensing}

\author{Hajrudin Bešić}\email{hajrudin.besic@tuwien.ac.at}% Your name
\affiliation{Institute of Sensor and Actuator Systems, TU Wien, Gusshausstrasse 27-29, 1040 Vienna, Austria}
\author{Alper Demir}
\affiliation{Department of Electrical Engineering, Koç University, Istanbul 34450, Turkey}
\author{Veljko Vukićević}
\affiliation{Institute of Sensor and Actuator Systems, TU Wien, Gusshausstrasse 27-29, 1040 Vienna, Austria}
\author{Johannes Steurer}
\affiliation{Institute of Sensor and Actuator Systems, TU Wien, Gusshausstrasse 27-29, 1040 Vienna, Austria}
\author{Silvan Schmid}
\affiliation{Institute of Sensor and Actuator Systems, TU Wien, Gusshausstrasse 27-29, 1040 Vienna, Austria}

\date{\today} % Leave empty to omit a date

\begin{abstract}
Nanomechanical sensors based on detecting and tracking resonance frequency shifts are to be used in many applications. Various open- and closed-loop tracking schemes, all offering a trade-off between speed and precision, have been studied both theoretically and experimentally. In this work, we advocate the use of a frequency counter as a frequency shift monitor in conjunction with a self-sustaining oscillator (SSO) nanoelectromechanical system (NEMS) configuration. We derive a theoretical model for characterizing the speed and precision of frequency measurements with state-of-the-art frequency counters. Based on the understanding provided by this model, we introduce novel enhancements to frequency counters that result in a trade-off characteristics which is on a par with the other tracking schemes. We describe a low-cost field-programmable-gate array (FPGA) based implementation for the proposed frequency counter and use it with the SSO-NEMS device in order to study its frequency tracking performance. We compare the proposed approach with the phase-locked-loop based scheme both in theory and experimentally. Our results show that similar or better performance can be achieved at a substantially lower cost and improved ease-of-use. We obtain almost perfect correspondence between the theoretical model predictions and the experimental measurements. 
\end{abstract}


\maketitle
\section{Introduction} \label{sec:introduction}

Frequency counters are a standard equipment to characterize the frequency fluctuations of oscillators and clocks, especially in estimating the well established time-domain measure of frequency stability, namely the Allan Deviation (AD)~\cite{dawkins2007considerations,benkler2015relation}. The averaging of instantaneous frequency over a certain observation (gate) time, which forms the basis for calculating AD, is naturally performed with a standard frequency counter. In this work, we propose using an improved frequency counter with high resolution and accuracy \cite{rubiola} as a frequency shift monitor for an oscillatory signal that is generated by an SSO-NEMS device, as opposed to simply using it as a tool for characterizing its raw frequency stability in the presence of thermomechanical and detection noise. The goal is to detect small frequency shifts due to events of interest, arising, e.g., from the interaction of the nanomechanical resonator with a mass, temperature, or force stimulus, as fast and precise as possible.  

We develop a theoretical model for characterizing the frequency counter measurements, and show that the averaging (gate) time of a standard counter can be used to balance the trade-off between the speed of detection and measurement precision. Based on the understanding provided by this model, we propose a novel counter architecture where the simple averaging of frequency over a gate time that spans across multiple signal cycles is replaced by a digital filter with adjustable bandwidth that operates on the resampled time stamps of the signal edges. The filtered time stamps are subsequently mapped to frequency measurements. We show that it is crucial to perform the filtering {\it before} the conversion of the time stamps to frequency values, especially in cases where transduction noise is dominant. While the proposed counter is not suitable for directly estimating the raw AD of the signal source anymore, it offers better trade-off characteristics as a frequency shift monitor. Conceptually, the output of the frequency counter could be used to synthesize a cleaner oscillatory signal that tracks the frequency shifts of interest but with subdued unwanted frequency fluctuations. We characterize the precision of the counter output by computing the AD of this conceptually synthesized signal.  

The standard and well established technique for tracking the frequency changes of an oscillatory signal source is a phase-locked loop (PLL), where the signal generated by a clean controlled-oscillator (CO) is phase- and frequency locked to the noisy signal source with a closed-loop feedback system \cite{best2007phase}. The feedback loop is designed so that the CO tracks the frequency shifts of interest while suppressing rapid fluctuations due to noise, with the loop bandwidth serving as the control knob for trading off tracking speed versus precision. The precision of the PLL output is characterized by computing the AD of the CO signal. In a PLL implementation, in addition to the CO, a phase difference (between the signal source and the CO output) detector is needed to generate the error signal in the feedback loop. In the context of NEMS based sensors, PLLs are usually realized using a lock-in amplifier based setup \cite{Demir,Pedram,hajrudinSSO}. 
Instead of locking a CO to the sensor signal to track its frequency, we recommend a new design using a frequency counter to directly measure the resonance frequency of the sensor. The sensor itself is excited by narrow pulses with low energy and oscillates freely \cite{hajrudinSSO}.
We use a reciprocal frequency counter in a continuous measurement mode where the counter hardware is not reset between measurements. This technique was first used in the HP 5371 frequency analyser \cite{hp_paper}. It greatly increases the number of samples. The use of continuous time interval measurements makes it easier to study the dynamic frequency behavior of a signal.

We compare the proposed self-sustaining oscillator with frequency counter scheme to the PLL approach both in theory and experiment, and show that similar or better performance can be achieved with respect to frequency resolution and stability of operation. We describe a low-cost FPGA based implementation of the proposed frequency counter. While the digital signal processing in a lock-in amplifier can also be implemented on an FPGA, considerably more resources are needed to implement the CO, the phase demodulators and the rest of the PLL functionality. Furthermore, only a low-Q bandpass filter is used to condition the signal for the frequency counter. Thus, the proposed frequency counter based scheme offers similar or better performance but at a substantially lower-cost and improved ease-of-use.

\section{Theory} \label{sec:theory}
\subsection{Interpolating Reciprocal Frequency Counter with Continuous Time Stamping}

We consider a state-of-the-art counter, namely an {\it interpolating reciprocal counter with continuous time stamping}~\cite{johansson2006new,rubiola}. In order to understand the speed and precision properties of such a counter used as a frequency shift monitor, we develop a simple model that captures its characteristics. Let $f_s(t)$ denote the instantaneous frequency (measured in units of Hz) of the signal source, which includes any fluctuations due to noise as well as shifts due to events of interest. We define $\phi(t)=\int^t f_s(u)\,du$ as the signal phase (unitless, equal to phase in radians divided by $2\pi$). In the frequency counter, time stamps for the boundaries of full signal cycles, {\it i.e.,} at the rising signal edges, are generated using a high-frequency, high-precision internal clock and an interpolator. That is, time $t_n$ where $\phi(t_n)=n$ ($n$ is an integer) is measured with a clock counter for full clock cycles and an interpolator between two clock edges that precede and succeed a signal edge~\cite{johansson2006new}. 

In the typical setting where a frequency counter is used to characterize the frequency stability of a high quality signal source, the resolution of the time stamps may be limited by the clock frequency and the quality of the interpolating circuitry. In the application we consider here, the signal source exhibits relatively large frequency fluctuations, resulting in deviations in the time stamps that are much larger than this resolution limit. In the model, we thus assume that time stamps $t_n$ can be measured precisely. In a reciprocal counter, the frequency of the source is estimated from the time stamps for one signal cycle with 
\begin{equation}
    f_{c}(t_n) = \frac{\int_{t_{n-1}}^{t_n}f_s(u)\,du}{t_n-t_{n-1}} = \frac{\phi(t_n)-\phi(t_{n-1})}{t_n-t_{n-1}} = \frac{1}{t_n-t_{n-1}}.
    \label{eqn:fcocycle}
\end{equation}
Thus, $f_{c}(t_n)$ represents the average of the instantaneous frequency $f_{s}(t)$ over one signal cycle between $t_{n-1}$ and $t_n$. The highest rate at which a frequency counter can generate an output is limited by the signal frequency (or  twice the signal frequency if falling signal edges are also used with a $50\%$ duty cycle). With (\ref{eqn:fcocycle}), the {\it gate time} of the counter is set to the cycle time of the signal source. A frequency estimate with gate time of $k$ cycles can be computed with 
\begin{equation}
    f_{c}^{(k)}(t_n) = \frac{k}{t_n-t_{n-k}}.
    \label{eqn:fcokcycle}
\end{equation}
If a sudden frequency shift occurs in the signal source between $t_{n-1}$ and $t_n$, its effect  will be fully reflected in the frequency estimate in (\ref{eqn:fcocycle}) at $t_{n+1}$ (within two cycles), whereas it will be at $t_{n+k}$ (within $k+1$ cycles) for the one in (\ref{eqn:fcokcycle}). However, the precision of the estimate in (\ref{eqn:fcokcycle}) is higher since rapid frequency fluctuations are suppressed due to the inherent averaging over $k$ cycles instead of just one. Thus, gate time can be used as a control knob for trading off response speed versus precision in a frequency counter that is used as a frequency shift monitor. 

The averaging inherent in (\ref{eqn:fcokcycle}) corresponds to a simple moving average filter. Instead, any filter that may offer a better response speed versus noise filtering characteristics can be used. In order to pursue this idea, we first need to better understand how frequency fluctuations affect the frequency estimates computed with a frequency counter. For a constant nominal signal frequency $f_o$, we consider
\begin{equation}
  \phi(t) = f_o\,(t+\alpha(t)),     
   \label{eqn:phwithnoise}
\end{equation}
where $\alpha(t)$ represents the time (phase) noise of the source.
Ideally, the instantaneous frequency $f_s(t)$, and the fractional frequency $y_s(t)$, can be computed from $\phi(t)$ with a time-derivative as follows 
\begin{equation}
f_s(t)=\dot{\phi}(t) = f_o\,(1 + \dot{\alpha}(t)) \;\;,\;\;  y_s(t)  = \frac{f_s (t)}{f_o} = 1 + \dot{\alpha}(t)
\label{eqn:phtofreqideal}
\end{equation}
where $\dot{\alpha}(t)$ represents the fractional frequency noise.

Let us now derive the fractional frequency estimate for a reciprocal counter. Based on (\ref{eqn:phwithnoise}), the time stamps $t_n$ satisfy $\phi(t_n) = n = f_o\,(t_n+\alpha(t_n))$, which yields $t_n = n\,T_o - \alpha(t_n)$ 
where $T_o=\tfrac{1}{f_o}$ is the nominal cycle time of the signal. We substitute this expression for $t_n$ in (\ref{eqn:fcocycle}) to derive the following
\begin{equation}
    y_c(t_n) = \frac{f_c(t_n)}{f_o} = \frac{1}{1-\frac{\alpha(t_n)-\alpha(t_{n-1})}{T_o}}
 \label{eqn:phtofreqreciprocal}   
\end{equation}
for the fractional frequency estimate computed in a reciprocal counter. The operation $\tfrac{\alpha(t_n)-\alpha(t_{n-1})}{T_o}$ in (\ref{eqn:phtofreqreciprocal}) above corresponds to the (discrete-time) derivative of $\alpha(t)$ over one cycle. Ideally, as in (\ref{eqn:phtofreqideal}), the conversion of phase to frequency is a linear transformation. However, in a reciprocal counter, this conversion involves a nonlinear operation, as seen in (\ref{eqn:phtofreqreciprocal}). If $z$  (fractional frequency noise) denotes the time derivative of $\alpha(t)$, the (ideal) linear transformation to fractional frequency can be represented by $1+z$ as in (\ref{eqn:phtofreqideal}), whereas it is given by the nonlinear function $\tfrac{1}{1-z}$ in (\ref{eqn:phtofreqreciprocal}) for a reciprocal counter. The power series expansion 
\begin{equation}
\frac{1}{1-z} = 1 + z + z^2 + \cdots    
\end{equation}
indicates that the two transformations are (approximately) equal only when $z$ is small. 

Detection noise (generated in the transduction of mechanical motion into an electrical signal) in a NEMS device results in white phase noise~\cite{Demir,Pedram,hajrudinSSO}. This corresponds to a frequency fluctuation spectrum that increases with frequency. Thus, $z$ may contain strong high frequency components. In this case, the quadratic $z^2$ term in the power series expansion needs to be taken into account to accurately characterize the fluctuations in the frequency estimated by a reciprocal counter. The high frequency spectral components in $z$ mix with each other through $z^2$ to produce low frequency fluctuations, known as {\it intermodulation noise}. In order to prevent or minimize the degrading impact of this nonlinear phenomenon on the accuracy of the frequency estimates computed by a frequency counter, a low-pass digital filter can be applied to the time stamps $t_n$, and hence to phase noise samples $\alpha(t_n)$, {\it before} the time stamps are converted to frequency estimates. Ideally when phase to frequency conversion is linear, a (linear) filter may be applied to the phase, or equivalently, to the frequency data, since the ordering of linear transformations does not change the final outcome. In the case of a reciprocal frequency counter, the low frequency intermodulation noise generated inherently in the conversion of time stamps to frequency estimates can not be removed with subsequent low-pass filtering. While time stamp to frequency conversion always generates intermodulation noise, its effect will be minimal if high frequency fluctuations are suppressed first, before the conversion, with a digital filter. The bandwidth and the characteristics of this filter can be chosen to trade off speed versus precision when the proposed counter is used as a frequency shift monitor.

\subsection{Sampling Rate and Decimation}
In a reciprocal counter, the sampling rate at the output is determined directly by the frequency of the input signal, given by $f_\mathrm{rate} = {f_\mathrm{s}}/{k}$, where $k$ is the number of cycles counted within one gate time. This input dependency results in problems  when subsequent digital signal processing is performed on the sampled frequency counter output. For instance, if any sort of filtering is performed, a change in the input signal frequency will consequently alter the sampling rate, thereby affecting the dynamics of the filter. To mitigate this issue, the input signal dependent sampling rate should be converted into a fixed one.

One approach to achieving a fixed sampling rate is to implement continuous event triggered timestamp counting as proposed in \cite{johansson2006new}. This involves using a dedicated counter that generates trigger events at regular intervals of $T_\mathrm{int}$, which determines the sampling rate. However, the samples cannot be taken precisely at multiples of  $T_\mathrm{int}$. Instead, they are generated at the next rising edge of the input signal. As a result, there is an inherent uncertainty in the sampling time, up to one period of the input signal. The impact of this uncertainty on the overall measurement varies depending on the number of signal periods encompassed within one interval. When there are numerous signal periods, the uncertainty has a lesser effect. However, if there are only a few signal periods, the irregular sampling interval introduces errors that can affect the accuracy of the measurement. 

Although the sampling rate ${f_\mathrm{s}}/{k}$ is directly linked to the input frequency $f_\mathrm{s}$, the sampling instants always align with the internal clock (with frequency $f_\mathrm{CLK}$) edges of the frequency counter. The sampling rate can be transformed up to $f_\mathrm{CLK}$ by simply interpolating the acquired data through the use of a zero-order hold. However, this introduces high-frequency harmonics due to the abrupt transitions between the samples. To address this, a low-pass filter can be applied to attenuate the harmonics. The combined process of low-pass filtering and decimation (to a fixed fraction of $f_\mathrm{CLK}$) after the zero-order hold can be efficiently achieved using a cascaded integrator-comb filter (CIC), as described in \cite{USigPro}.
\begin{figure}
    \centering
    \captionsetup[subfigure]{position=top,singlelinecheck=off,justification=raggedright}
    \subfloat[]{\includegraphics[width=\columnwidth]{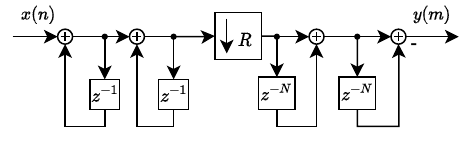}\label{fig:cic_block}}
        
    \subfloat[]{\includegraphics[width=\columnwidth]{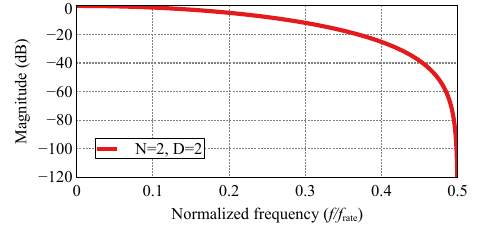}\label{fig:cic_tf}}
    \caption{a) Block diagram of the second order CIC decimator with the decimation factor $R$ and the comb section delay $N$. b) Transfer function of the decimated output.
    }
\end{figure}
Figure \ref{fig:cic_block} depicts the second-order CIC filter employed in this study, featuring two integrator sections and two comb sections. The comb sections introduce a delay of $N=2$. The downsampler, with a value of $R=2^{13}$, is positioned between the integration and comb sections. Thus, the sampling rate of the final output is given by $f_\mathrm{CLK}/R$, independent of the input frequency $f_\mathrm{s}$. Figure \ref{fig:cic_tf} shows the transfer function of the CIC filter after decimation. Notably, it does not exhibit a distinct separation between the pass-band and stop-band, thus necessitating further filtering with a finite impulse response (FIR) or infinite impulse response (IIR) filter.

\subsection{Allan Deviation}

Allan Deviation (AD)  $\sigma_y(\tau)$ is a widely used and well-established method for characterizing frequency fluctuations \cite{silvanbook2nd, Demir, Allan_Paper}. For AD, the frequency values need to be normalized, resulting in a fractional frequency
\begin{equation}
    y(t)=\frac{\Delta\omega(t)}{\omega_0}.
\end{equation}
AD is  the square root of the Allan Variance, which can be computed from sampled frequency data using the following equation
\begin{equation}
    \sigma_y^2(\tau) =\frac{1}{2\,(N-1)} \sum_{i=1}^N (y_{i+1,\tau}-y_{i,\tau})^2.
    \label{eq:avar1}%
\end{equation}
Here, $y_i$ represents the $i\,{\mathrm{th}}$ sample of the {\em averaged} frequency over the averaging time $\tau$. It is calculated as
\begin{equation}
    y_{i,\tau} = \frac{1}{\tau}\int_{(i-1)\tau}^{i \tau}y(t)\mathrm{d}t.
    \label{eq:avar2}%
\end{equation}
Thus, the averaging operation above has to be part of the frequency measurement and data acquisition process. A frequency counter naturally performs this operation when it is used to characterize the raw frequency fluctuations of its input signal. On the other hand, equation \eqref{eq:avar1} is routinely used in practice on measured frequency data, also in cases where a frequency counter is not used and/or the averaging operation is not inherently included in the measurement process. Such use is justified only when the sampling rate is sufficiently large when compared with the system bandwidth, i.e., the smallest time $\tau$ that determines the sampling interval is small enough. In this case, the frequency that is being measured is (almost) constant over the sampling interval (smallest $\tau$) resulting in 
\begin{equation}
    y_{i,\tau} = \frac{1}{\tau}\int_{(i-1)\tau}^{i \tau}y(t)\mathrm{d}t \approx y(i \tau).
    \label{eq:avar3}%
\end{equation}
Averaging when $\tau$ is equal to an integer $m$ multiple of the sampling interval is performed by simply averaging $m$ consecutive samples of the acquired raw data.  

When a frequency counter is employed as a frequency shift monitor, as we propose in this work, in contrast with its use in characterizing the frequency fluctuations of its {\em input} signal, AD should be computed for the conceptually synthesized oscillatory signal based on the counter {\em output}. Thus, the averaging in \eqref{eq:avar2} needs to be performed {\em in addition} to the inherent averaging performed by the counter itself. While the inherent averaging of the counter is over a time interval determined by its gate time specification, the average in \eqref{eq:avar2} is computed for all $\tau$ that is of interest in the AD characterization.   

The Allan Variance can alternatively be computed in the frequency domain if the power spectral density of fractional frequency fluctuations, denoted by $S_y(\omega)$, is known. The equation for this computation, shown below,
\begin{equation}
    \sigma_\mathrm{y}^2(\tau)= \frac{1}{2\pi}\frac{8}{\tau^2} \int_0^\infty \frac{[\sin(\frac{\omega\tau}{2})]^4}{\omega^2} S_\mathrm{y}(\omega)\mathrm{d}\omega
    \label{eq:sigma}
\end{equation}
includes the frequency domain equivalent of the averaging operation in \eqref{eq:avar2}, should thus be used on $S_y(\omega)$ which was measured or computed without inherent averaging.
For white frequency fluctuations, where $S_y(\omega)$ is constant, equation (\ref{eq:sigma}) simplifies to
\begin{equation}\label{eq:one-over-sqrttau}
    \sigma_y^2=\frac{S_y(0)}{2 \tau}.
\end{equation}
Hence, in systems limited by thermal white noise, the resulting AD exhibits a $\sigma_y\propto 1/\sqrt{\tau}$ dependence on the averaging time $\tau$.

This study focuses on two primary noise sources, thermomechanical noise and detection noise. Thermomechanical noise is regarded as the fundamental noise source in NEMS resonators, resulting from the random movement of resonator molecules. On the other hand, detection noise arises from the transduction of the resonator's mechanical motion into an electrical signal, as well as from electronic components involved in the detection process. 

The power spectral density of frequency noise, $S_{\Delta \omega}(\omega)$, can be computed as a superposition of the power spectral densities of thermomechanical and detection phase noise, multiplied by their corresponding transfer functions (magnitude squared) as derived in \cite{Demir,hajrudinSSO}. This yields the spectral density of fractional frequency noise required for computing the AD
\begin{equation}
\begin{split}
    S_\mathrm{y}(\omega)&= \frac{S_{\Delta \omega}(\omega)}{\omega_0^2}\\
    &= \frac{S_{\theta_\mathrm{th}}(\omega)|H_{\theta_\mathrm{th}}(\mathrm{j}\omega)|^2+ S_{\theta_\mathrm{d}}(\omega)|H_{\theta_\mathrm{d}}(\mathrm{j}\omega)|^2}{\omega_0^2}\\
    &= \frac{S_{\theta_\mathrm{th}}(\omega)}{\omega_0^2}\left[|H_{\theta_\mathrm{th}}(\mathrm{j}\omega)|^2 + \mathcal{K}^2|H_{\theta_\mathrm{d}}(\mathrm{j}\omega)|^2 \right],
\end{split}
\label{eq:S(omega)}
\end{equation}
where $\mathcal{K} = \sqrt{{S_{\theta_\mathrm{d}}}/{S_{\theta_\mathrm{th}}}}$ is the ratio between the thermal and detection noise and $H_{\theta_\mathrm{th}}$ and $H_{\theta_\mathrm{d}}$ are the transfer function of the thermomechanical and detection noise to the frequency output, respectively. For the SSO frequency tracking scheme, the transfer functions are given by \cite{hajrudinSSO}
\begin{equation}
\begin{split}
    H_\mathrm{\theta_\mathrm{th}}^\mathrm{SSO}(s) = \frac{1}{\tau_\mathrm{r}}H_\mathrm{L}(s),\\
    H_\mathrm{\theta_\mathrm{d}}^\mathrm{SSO}(s) = \frac{1}{\tau_\mathrm{r}}\frac{1}{H_\mathrm{r}(s)} H_\mathrm{L}(s),
\end{split}
\label{eq:sso_tf}
\end{equation}
where $H_\mathrm{r}$ is a single-pole low-pass filter with the time constant of the resonator. $H_\mathrm{L}$ has low-pass characteristics and represents the bandwidth limiting (noise filtering) mechanisms in the frequency detection device. The theoretical computation of AD is performed with the frequency domain approach in  \eqref{eq:sigma}, where the spectral density of fractional frequency noise is first computed using a frequency domain model of the SSO and the frequency counter, as in \eqref{eq:S(omega)} and \eqref{eq:sso_tf}, that includes intermodulation noise generated in the time stamp to frequency conversion.  

\section{Methods} \label{sec:methods}

We describe the experimental setup for the comparison of the two methods, the proposed frequency counter and the PLL frequency detector (PLLFD), for frequency shift monitoring of an oscillator. Our measure of assessment for precision is the Allan Deviation (AD). Central to our investigation is the self-sustaining oscillator, as detailed in \cite{hajrudinSSO}, driven by narrow pulses and oscillating freely. The pulse duration and timing is automatically adjusted to attain the desired resonance frequency within a closed-loop setup involving the resonator and the pulse generation mechanisms. Notably, the frequency measurement and detection operate outside of this loop. Our study involves resonance frequency measurements using the proposed frequency counter and the established PLLFD, as depicted in Figure \ref{fig:SSO_block}. We compute the AD in two experimental conditions, using the frequency counter output and the PLLFD output. For the frequency counter, we explore filter configurations, cut-off frequencies, and timestamp versus frequency filtering.

\subsection{NEMS Resonator Setup}
In this study, we utilized a NEMS resonator consisting of a square \SI{50}{\nano\meter} thick silicon nitride membrane measuring \SI{1018}{\micro\meter} on each side, as introduced in \cite{hajrudinSSO}. To achieve electrical transduction, we incorporated two \SI{5}{\micro\meter}-wide gold (Au) electrodes spanning over the resonator. The membrane was placed within a static magnetic field of approximately \SI{0.8}{\tesla}, generated by a Halbach array composed of neodymium magnets. The orientation of the traces was perpendicular to the magnetic field. By capitalizing on the resulting Lorentz force, one metal trace served for the purpose of driving the resonator with an AC current and the second electrode for detecting its motion through the magnetomotively induced voltage.

To amplify the detected signal from the metal trace, we employed a custom-made, low-noise differential pre-amplifier with a gain factor of $10^4$. The NEMS was operated in vacuum with a pressure of $8.2\cdot10^{-6}\,\mathrm{mbar}$. The NEMS resonator had a resonance frequency of $\omega_r=119$~kHz and a quality factor of $Q = 57.5\,\mathrm{k}$. Consequently, the response time of the NEMS resonator is calculated with $\tau_\mathrm{r}= 2Q/\omega_r$, is $154$~ms.

\subsection{Self-Sustaining Oscillator}
\begin{figure}
    \centering
    \includegraphics[width=\columnwidth]{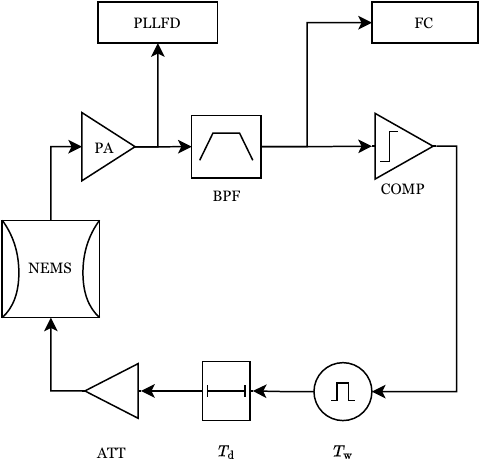}
    \caption{Block representation of the self-sustaining oscillator tracking scheme with the frequency counter (FC) and the phase-locked loop frequency detector (PLLFD).}
    \label{fig:SSO_block}
\end{figure}

The resonator utilized in this study operates as a self-sustaining oscillator (implemented in PHILL from Invisible-Light Labs GmbH), as described in \cite{hajrudinSSO} and depicted in Figure~\ref{fig:SSO_block}. The transduced output of the NEMS resonator is connected to a preamplifier (PA). The amplified signal is then directed to a band-pass filter (BPF), serving two purposes, to suppress unwanted modes and to restrict the noise bandwidth of the system to prevent noise folding (aliasing) in the frequency counter, as it will be discussed in detail. %On the other hand, the bandwidth of the band-pass filter should be large enough to ensure that the response speed of the system is not limited.
The output of the band-pass filter is linked to a comparator (COMP), which transforms the sinusoidal signal into a rectangular waveform, triggering the pulse generation mechanism that drives the NEMS resonator. The pulse generation mechanism is comprised of two components: One generates a pulse with a width of $T_\mathrm{w}$, while the other delays the pulse generated at the feedback output by a time of $T_\mathrm{d}$. Since the NEMS resonator can only tolerate low currents, the generated pulse needs to be attenuated (ATT) by a factor of $10^5$ before being applied to the NEMS.

Frequency detection is accomplished using two different methods, the PLLFD and our proposed frequency counter (FC). The PLLFD is connected to the output of the preamplifier, while the FC is connected to the output of the band-pass filter.

\begin{figure*}
    \centering
    \includegraphics{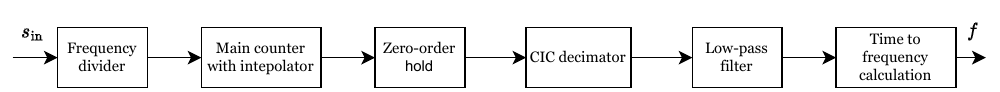}
    \caption{Block diagram of the proposed frequency counter architecture, where $s_\mathrm{in}$ is the input signal, $f$ is the frequency output.}
    \label{fig:fc_block}
\end{figure*}

\subsection{Frequency Counter}
Figure \ref{fig:fc_block} illustrates the block diagram of the frequency counter architecture. The frequency counter employed in this study is an enhanced version of a state-of-the-art interpolating reciprocal counter with continuous trigger events (implemented in PHILL from Invisible-Light Labs GmbH). It has been modified to better suit the task of tracking resonance frequency changes with improved usability. 

The front-end of the frequency counter is a frequency divider, which enables counting $k$ cycles of the input signal, where $k$ can be any integer and specifies the number of counted periods in one timestamp. To avoid aliasing and noise folding, it is recommended to keep $k$ as small as possible, i.e., set it to $k=1$, since the sampling rate of the counter, $f_\mathrm{rate}$, is directly dependent on the input signal frequency, given by $f_\mathrm{rate}={f_\mathrm{s}}/{k}$.

The subsequent stage of the system is the main counter with an interpolator. It functions as a standard reciprocal counter, tallying the number of rising edges of the internal clock that occur in an interval of $k$ periods of the input signal and generating the time stamps for the interval boundaries. However, due to the potential error of up to one clock cycle in such a counter, an interpolation mechanism is employed to enhance the precision of the time stamps up to \SI{100}{\pico\second} at \SI{100}{\kilo\hertz}.

The output rate of the main counter is directly linked to the frequency of its input, necessitating resampling to achieve a fixed sampling rate. First, the time stamp series is interpolated onto the internal clock transitions time grid using a zero-order hold. Second, it is decimated to the desired sampling rate through a CIC decimator. Third, the resampled time stamp series is low-pass filtered, which defines the final system bandwidth as specified by $H_\mathrm{L}$ in equation \eqref{eq:sso_tf}. Finally, the filtered time stamp series is converted to the frequency output using equation \eqref{eqn:fcocycle} or \eqref{eqn:fcokcycle}.

\section{Results and Discussion} \label{sec:results}
We conducted ten-second measurements to analyze the output frequency fluctuations due to thermomechanical and detection noise for various frequency shift detection scenarios and system parameter choices. The experimental setup, which incorporated magneto-motive readout, introduced significant detection noise. Consequently, the thermomechanical noise peak could not be resolved at the resonator output, resulting in a value of $\mathcal{K}> 1$. As discussed in \cite{hajrudinSSO,silvanbook2nd}, this leads to an AD with a $1/\tau$ dependency if the system time constant is shorter than the resonator time constant.
In the following, we examine the frequency counter under four different conditions.

\subsection{Filtering of time stamp data versus frequency data}
\begin{figure}
    \centering
    \includegraphics{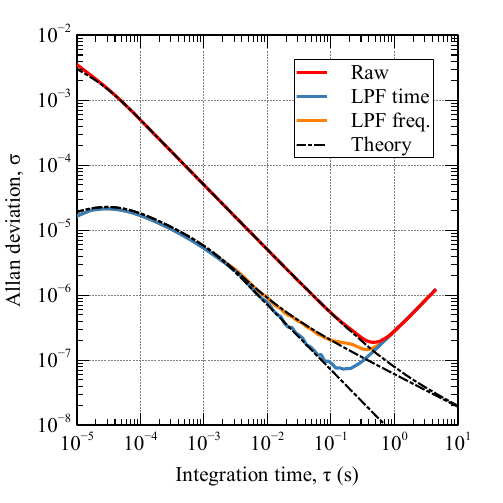}
    \caption{Effect of the position (before or after the time to frequency conversion) of a first-order low-pass filter with a cut-off frequency of \SI{200}{\hertz} on the AD. An SSO loop BPF with a \SI{20}{\kilo\hertz} bandwidth was used for all experiments.}
    \label{fig:time_vs_freq_filter}
\end{figure}

In the first experiment, shown in Figure~\ref{fig:time_vs_freq_filter}, we compare two alternatives for a frequency counter, with conversion of time stamp series to frequency data {\em after} or {\em before} low-pass filtering. As articulated in Section~\ref{sec:theory}, the expected $1/\tau$ dependence of the AD is altered to $1/\sqrt{\tau}$ for large $\tau$. This is due to the mixing of the high-frequency noise components by the non-linearity of the time-to-frequency conversion process, resulting in intermodulation noise at lower frequencies. This can be alleviated by removing or reducing high-frequency noise before time-to-frequency conversion by low-pass filtering the counter output in the form of a time stamp series. %The latter approach is more practical and flexible in real-world applications as using a narrow band-pass filter limits the system dynamics and response time as dictated by the filter. Alternatively, employing a wide band-pass filter and subsequently applying a low-pass filter to the counter output provides the most flexibility for operation within a broader range of system bandwidth. 
The ADs of these measurements are shown in Figure~\ref{fig:time_vs_freq_filter} showing that filtering the time stamp data first and then conversion to frequency yields better AD than conversion before filtering, as predicted by theory. %The band-pass filter was present in the generation of all of the measurements shown in Figure \ref{fig:time_vs_freq_filter}, whereas the low-pass filter applies only to the data labeled as LPF. 

\subsection{Sweeping the gate time of the counter}
\begin{figure}
    \centering
    \includegraphics[width=\columnwidth]{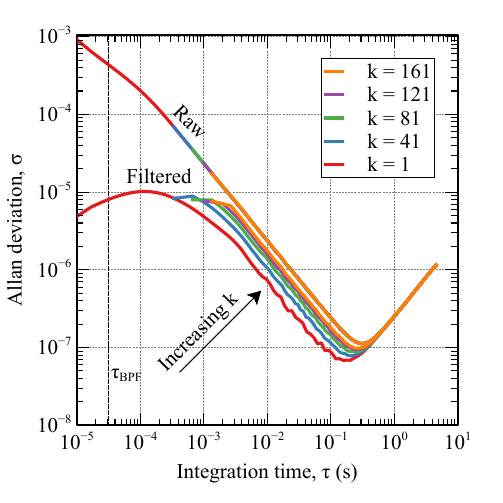}
    \caption{Comparison between the ADs of raw and filtered (in the frequency counter with a first order low-pass filter with a cut-off frequency of \SI{200}{\hertz}) data for varying gate time as set with $k$ counted periods. The SSO loop band-pass filter (BPF) used in this experiment had a bandwidth of \SI{5}{\kilo \hertz} with the corresponding time constant $\tau_{BPF}$.}
    \label{fig:ps_filtering}
\end{figure}

In the second experiment, we investigate the impact of varying the number of counted periods $k$ of the signal, i.e., the gate time of the counter, on the performance and the filtering process in the frequency counter. As seen in Figure~\ref{fig:ps_filtering}, increasing $k$ seemingly does not lead to an improvement in the AD without suppressing it (raw). In contrast, low-pass filtering the frequency counter output results in improved ADs. The best AD is observed for $k=1$ with the low-pass filter.

ADs for the unfiltered raw data at the counter output fall exactly on top of each other when $k$ is varied, albeit starting at larger $\tau$ values for larger $k$ due to the larger sampling interval of $k$ periods. This result is puzzling at first thought, since theoretical considerations indicate that we should be able to use the gate time of the counter as a control knob for trading off response speed versus precision. While the response time of the counter is definitely prolonged with increased gate time (sudden frequency shift at the input will be fully reflected in the counter output after $k$ periods of the input, as discussed before), results in Figure \ref{fig:ps_filtering} suggest that there is no improvement in the precision of the output.  

The seemingly unexpected result in Figure \ref{fig:ps_filtering} can be resolved and deciphered as follows. The sampling rate at the output of the counter is inversely proportional to the gate time, with the counter producing a measurement for every $k$ periods of the input signal. When AD is computed based on this sampled frequency data for the smallest $\tau$ of $k$ periods, the only averaging in the sense of equation \eqref{eq:avar2} reflected in the data is the one that is inherently performed by the counter front-end, with no additional averaging in the actual AD computation as discussed before. This means that the AD computed as such is actually a characterization of the frequency fluctuations of the counter input signal, as opposed to the conceptually synthesized signal based on the counter output. Thus, it makes perfect sense that computed ADs for varying $k$ fall exactly on top of each other, since they all represent a characterization of the input signal, not the counter output. Another perspective on the seemingly puzzling results in Figure~\ref{fig:ps_filtering} is as follows. With increased gate time, the sampling rate at the output of the counter is not large enough to justify the use of the frequency data to compute the AD (for the counter output) with equation \eqref{eq:avar1}, since equation \eqref{eq:avar3} does not hold in this case, and the additional averaging that needs to be part of the AD computation is missing.      

% AD: Removed below content due to the new discussion above.
% The reason behind this lies in the similarity between the acquisition process of the counter (equation \ref{eqn:fcocycle}) and the calculation of the Allan deviation (equation \ref{eq:avar2}). Both methods involve averaging over a defined time period ($\tau$), lacking any memory or dynamic coupling to previous samples. Increasing the gate time (i.e., the number of counted signal periods) solely extends the averaging time for a single sample. While this does slow down the average step response speed of the counter, it fails to suppress detection noise to the same extent as a low-pass filter. Low-pass filters possess dynamic states that retain information from previous samples.

\begin{figure}
    \centering
    \includegraphics{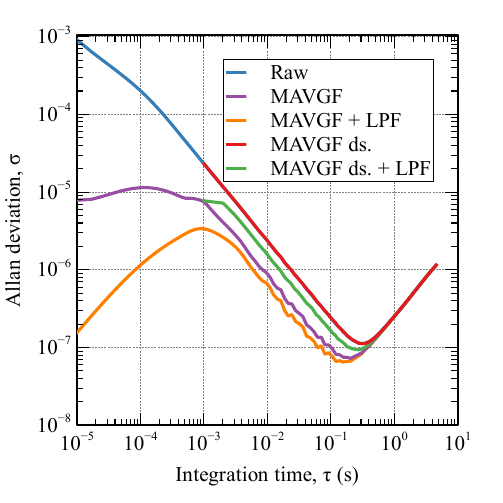}
    \caption{Measurement-based model of the frequency counter acquisition mechanism emulating a large gate time with $k=121$ by applying a moving average filter and downsampling to data acquired with $k=1$ and comparison of low-pass filtering at different stages. (LPF - low-pass filter)}
    \label{fig:fcmavg}
\end{figure}

While we resolved the results in Figure \ref{fig:ps_filtering} based on theoretical arguments, it is desirable to experimentally observe the precision improvement one can obtain in a frequency counter by increasing the gate time. This requires somehow increasing the sampling rate at the counter output. With a gate time of $k$ input periods, one can attain a $k$-fold increased sampling rate by employing a sequence of $k$ parallel counter front-ends, where each front-end counts over $k$ periods of the input signal, but with one period delay relative to the previous one in an interleaved manner. However, this would increase the hardware cost and complexity considerably. Instead, we simply emulate $k$ parallel counter front-ends by first producing output from the counter with $k=1$ for every period of the signal, and then processing this output with a {\em moving average filter} (MAVGF) with a window length of $k$. Apart from a larger latency at the output, this emulation is hardware-equivalent to having $k$ parallel counter front-ends and does not involve any approximations. One can obtain the frequency data at the $k$-fold lower sampling rate (output of only one of the $k$ front-ends, used to generate Figure \ref{fig:ps_filtering}) by simply downsampling the moving average filter output with a downsampling ratio of $k$. 

The ADs computed from the raw counter output with $k=1$, the moving average filter output with $k=121$, the downsampled moving average output, as well as low-pass filter processed versions of the moving average output and its downsampled version, are shown in Figure \ref{fig:fcmavg}. As expected, we observe the perfect coincidence of the AD curve for the downsampled moving average output with the curve for the raw counter output. 
We now also observe the theoretically claimed precision improvement at the counter output with increased gate time (emulated with the moving average output) even when there is no extra low-pass filtering. With increased sampling rate, computation of AD with equation \eqref{eq:avar1} is justified since equation \eqref{eq:avar3} now holds. The additional averaging required in AD computation for $\tau$ values larger than the sampling interval is performed using the data samples available at the higher rate. 

Further low-pass filtering the downsampled moving average output results in a precision improvement as indicated by Figure \ref{fig:fcmavg}, but not as much as the one we observe for the moving average output at the higher sampling rate and its filtered version. On the other hand, it seems strange that the moving average output has higher precision when compared with its downsampled version. In the end, specific frequency measurement values in the downsampled data are {\em exactly equal} to a subset of the values in the higher rate data, and therefore should have the same precision. In fact, use of AD in order to characterize the precision of the downsampled data is not appropriate since equation \eqref{eq:avar3} does not hold. However, one can simply compute the {\em standard} deviation of the downsampled data, which is in fact {\em equal} to the standard deviation of the higher-rate date assuming statistical stationarity. Any digital signal processing with memory, that involves dynamics over time (including AD computation), on the downsampled data is not meaningful, and introduces aliasing and noise folding due to the low sampling rate. When gate time is set to the lowest value, the input signal period with $k=1$, the counter front-end averages over the signal period $T_\mathrm{s}$, which is also set to the sampling interval at the counter output. Therefore, a front-end band-pass filter with a maximum bandwidth set to half the signal frequency is needed to prevent aliasing and noise folding, even when $k=1$.

We thus conclude that, even though gate time can be used as a control knob for trading off precision versus response speed when a frequency counter is used as a frequency shift monitor, it is better to set the gate time to the smallest value with $k=1$ and generate output at the largest rate possible, with an appropriate front-end band-pass filter to prevent aliasing. Then, instead of emulating a larger gate time with a simple moving average filter (a specific type of FIR filter), it is better to use an appropriate, bandwidth adjustable FIR or IIR digital filter that can offer a better precision versus speed trade-off. For the results we present in this work, we used a first-order IIR low-pass filter.

\subsection{Resampling for a fixed sampling rate}

\begin{figure}
    \centering
    \includegraphics[width=\columnwidth]{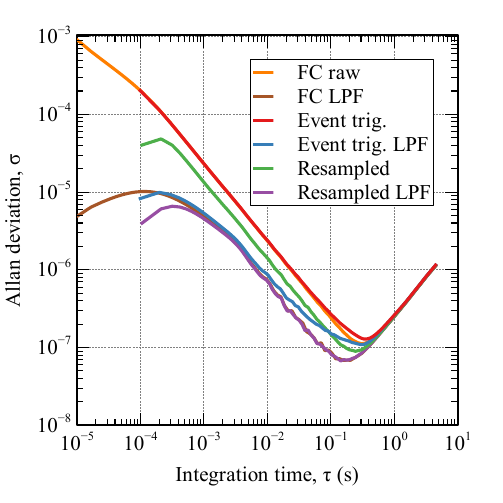}
    \caption{ADs showing that resampling and filtering the signal at a fixed sampling rate achieves the same performance, whereas the event triggered timestamp method results in a degradation. (LPF - low-pass filter)}
    \label{fig:direct_vs_dec}
\end{figure}

In the third experiment, we consider resampling of the frequency counter output, which has an input dependent sampling rate, to a fixed sampling frequency. We compare filtering of the raw output and the resampled version. We also compare our proposed resampling technique with the continuous event triggered timestamp method described in \cite{johansson2006new}.  

We consider and compute six versions of the frequency data: (i) The raw main counter output (for $k=1$) with an input dependent sampling rate is generated. (ii) The raw output  is passed through a first-order Butterworth low-pass filter (LPF) with a cut-off frequency of \SI{200}{\hertz}. (iii) The raw output is processed using the event triggered timestamp method with $T_\mathrm{int} = \SI{100}{\micro \second}$. (iv) The output processed with the event triggered timestamp method is also passed through the LPF. (v) The raw output is first processed with a zero-order hold to increase the sampling frequency to the frequency $f_\mathrm{CLK}=76.92 \,\unit{MHz}$ of the internal clock. Subsequently, it is decimated using a CIC decimator with a decimation factor of $R=8192$, resulting in a final sampling rate of $9.4 \,\unit{kHz}$. (vi) The resampled data is passed through the LPF. We note that the digital LPF is implemented at the respective sampling rate of the data it is applied to, corresponding to the same cut-off frequency in each case.   

% AD: Streamlined the below description in the paragraph above. 
% The signal obtained from the main counter is initially filtered using a first-order Butterworth low-pass filter with a cut-off frequency of \SI{200}{\hertz}. The timestamp signal is obtained by sampling at rising edges of the signal right after a timestamp event occurs. The period of the timestamp events is set to \SI{10}{\micro \second}. On the other hand, the resampled signal with a fixed sampling rate is generated by applying a zero-order hold to the output of the main counter and placing it on a grid with a sampling frequency of $f_\mathrm{rate}=f_\mathrm{clk}=76.92 \,\unit{MHz}$. Subsequently, it is decimated using a CIC decimator with a decimation factor of $R=8192$, resulting in a sampling rate of $f_\mathrm{rate}=9.4 \,\unit{kHz}$. Following the decimation, the signal is filtered by a low-pass filter with a cut-off frequency of \SI{200}{\hertz}.

The ADs for the six versions of frequency data described above are shown in Figure \ref{fig:direct_vs_dec}. The event triggered timestamp method produces an uncertainty in the sampling instants up to one period of the input signal, which manifests itself in the form of additional frequency noise resulting in a larger AD, discernible in Figure \ref{fig:direct_vs_dec} right before thermal drift kicks in. The additional frequency noise produced by this technique, which cannot be suppressed,  is even more noticeable after the data is processed with the LPF. On the other hand, the data produced by our proposed resampling technique exhibits a slightly improved AD compared to the raw main counter output. This improvement can be attributed to the inherent low-pass filtering performed by the CIC decimator. Furthermore, if the resampled signal is further processed with an LPF,  the AD obtained is identical to that when the raw counter output is low-pass filtered with the same cut-off frequency.

\begin{figure}
    \centering
    \includegraphics[width=\columnwidth]{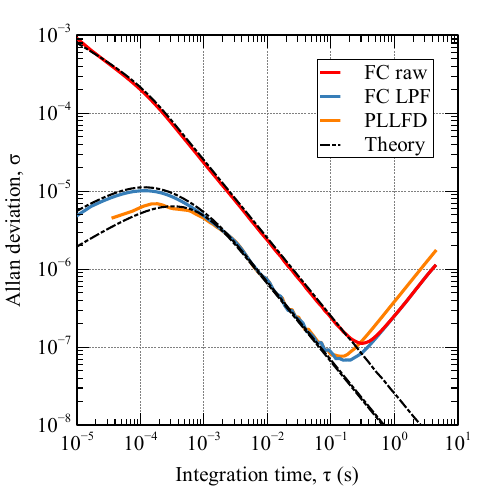}
    \caption{ADs of raw and low-pass filtered frequency counter output in comparison with a PLLFD with the same bandwidth. Results show that there is no difference in performance between a commercial PLLFD with the frequency counter proposed in this work.}
    \label{fig:pll_vs_fc}
\end{figure}

\subsection{Proposed frequency counter versus PLLFD}

In the final experiment, we compare the proposed frequency counter with a commercial, lock-in based PLLFD. The PID (Proportional-Integral-Derivative) coefficients of the PLLFD are generated by the software that comes with the equipment, targeting a loop bandwidth of \SI{200}{\hertz}, resulting in the values $k_\mathrm{p}=\SI{2.92}{\hertz/\deg}$ and $k_\mathrm{i}=\SI{13.34}{\hertz/\deg/\sec}$. The low-pass filter in the PLL demodulator is a first-order filter with a cut-off frequency of \SI{1}{\kilo \hertz}, and the PLL operates at a sampling rate of \SI{27}{\kilo \hertz}. The transduced NEMS output after the pre-amplifier is fed to the PLLFD, whereas it is processed with a band-pass filter (with a bandwidth of \SI{5}{\kilo \hertz}) to produce the input for the frequency counter. The frequency counter raw output (with $k=1$) is processed with a CIC decimator with a decimation factor of $R=8192$, resulting in a final sampling rate of \SI{9.4}{\kilo \hertz}. The output of the decimator is processed with a low-pass filter with a cut-off frequency of \SI{200}{\hertz}.

Figure \ref{fig:pll_vs_fc} illustrates the comparison between the state-of-the-art PLLFD method for frequency tracking and the proposed frequency counter based technique. It can be observed that both methods exhibit almost the same performance in both measurement and theory.

\section{Conclusions} \label{sec:conclusions}

In this study, we investigated various aspects of frequency shift monitoring mechanisms based on frequency counters for resonant sensors. We characterized their precision, both in theory and experimentally, in the presence of thermomechanical and detection noise. Through theoretical models, analyses and articulate arguments, combined with a series of experiments and elucidated results, we have gained valuable insights into various scenarios, system architecture and parameter choices. 

We proposed a novel and cost-effective frequency counter based frequency shift monitoring scheme, which was overlooked in the NEMS literature. Our frequency counter based architecture features wide band-pass filtering for signal conditioning combined with digital low-pass filtering in the sampled data domain before time stamp series for the signal transitions are converted to frequency data. This architecture not only alleviates the detrimental effect of intermodulation noise generated by the non-linearity of time-to-frequency conversion, but also enables a flexible and practical platform for real-world applications where all digital signal processing is performed at a fixed, input independent sampling rate.  

We investigated mechanisms for trading-off response speed versus precision in our proposed frequency counter based scheme. Although we have shown that the gate time of a counter can be used as a control knob for this purpose, it is more effective and efficient if the gate time is set to the smallest possible value, i.e., the cycle time of the input signal, to generate an output at the largest sampling rate possible. Then, speed versus precision trade-off can be conveniently achieved by modifying the digital filtering characteristics in a more flexible manner.  
% AD: Removed below content due to new conclusions above. 
% Next, we explored the influence of the number of counted periods on the filtering process. Contrary to expectations, increasing the number of counted periods did not lead to an improvement in the Allan deviation. Instead, it resulted in the loss of information about higher frequencies without effectively suppressing the deviation slope. This can be attributed to the lack of memory and dynamic coupling in the averaging process of the counter, which is not as effective in noise suppression as a low-pass filter with its inherent ability to retain information from previous samples.

The output of a standard reciprocal counter, where gate time is set to the cycle time of the input signal, has an input dependent sampling rate. This is inconvenient for the subsequent digital signal processing. We developed a resampling scheme that involves a zero-order hold and a CIC decimator to convert the output to a fixed, input independent sampling rate. We have shown experimentally that the resampled and filtered output has the same (or slightly better) precision as the filtered raw output of the frequency counter as characterized by ADs.  

% Furthermore, we investigated the benefits of fixing the sampling rate and filtering compared to direct filtering of the raw frequency counter output. The signal, obtained by resampling at a fixed sampling frequency and subsequent decimation, exhibited a slightly improved Allan deviation due to the inherent filtering characteristics of the decimator. Additional filtering of the decimated signal yielded the same Allan deviation as filtering the raw output from the frequency counter, demonstrating the effectiveness of fixing the sampling rate.

Finally, we compared the precision of the proposed frequency counter based scheme with a commercial implementation of the common and standard PLLFD technique in terms of ADs. Our results showed that both methods achieve the same performance, indicating that the proposed frequency counter based frequency tracking method can serve as a viable and cost-effective alternative to the state-of-the-art PLLFD method.

Overall, our experimental results, theoretical models, analyses and findings contribute to a better understanding of frequency detection mechanisms. This understanding combined with our FPGA based flexible and cost-effective platform paves the way to developing new frequency shift monitoring schemes with enhanced precision, response speed and/or reliability.  

\section*{Acknowledgements} \label{sec:acknowledgements}
This project received funding from the European Innovation Council under the European Union's Horizon Europe Transition Open program (grant agreement-101058711-NEMILIES).
\bibliographystyle{apsrev4-1}
\bibliography{bibliography.bib}

\begin{thebibliography}{12}%
\makeatletter
\providecommand \@ifxundefined [1]{%
 \@ifx{#1\undefined}
}%
\providecommand \@ifnum [1]{%
 \ifnum #1\expandafter \@firstoftwo
 \else \expandafter \@secondoftwo
 \fi
}%
\providecommand \@ifx [1]{%
 \ifx #1\expandafter \@firstoftwo
 \else \expandafter \@secondoftwo
 \fi
}%
\providecommand \natexlab [1]{#1}%
\providecommand \enquote  [1]{``#1''}%
\providecommand \bibnamefont  [1]{#1}%
\providecommand \bibfnamefont [1]{#1}%
\providecommand \citenamefont [1]{#1}%
\providecommand \href@noop [0]{\@secondoftwo}%
\providecommand \href [0]{\begingroup \@sanitize@url \@href}%
\providecommand \@href[1]{\@@startlink{#1}\@@href}%
\providecommand \@@href[1]{\endgroup#1\@@endlink}%
\providecommand \@sanitize@url [0]{\catcode `\\12\catcode `\$12\catcode
  `\&12\catcode `\#12\catcode `\^12\catcode `\_12\catcode `\%12\relax}%
\providecommand \@@startlink[1]{}%
\providecommand \@@endlink[0]{}%
\providecommand \url  [0]{\begingroup\@sanitize@url \@url }%
\providecommand \@url [1]{\endgroup\@href {#1}{\urlprefix }}%
\providecommand \urlprefix  [0]{URL }%
\providecommand \Eprint [0]{\href }%
\providecommand \doibase [0]{http://dx.doi.org/}%
\providecommand \selectlanguage [0]{\@gobble}%
\providecommand \bibinfo  [0]{\@secondoftwo}%
\providecommand \bibfield  [0]{\@secondoftwo}%
\providecommand \translation [1]{[#1]}%
\providecommand \BibitemOpen [0]{}%
\providecommand \bibitemStop [0]{}%
\providecommand \bibitemNoStop [0]{.\EOS\space}%
\providecommand \EOS [0]{\spacefactor3000\relax}%
\providecommand \BibitemShut  [1]{\csname bibitem#1\endcsname}%
\let\auto@bib@innerbib\@empty
%</preamble>
\bibitem [{\citenamefont {Dawkins}\ \emph {et~al.}(2007)\citenamefont
  {Dawkins}, \citenamefont {McFerran},\ and\ \citenamefont
  {Luiten}}]{dawkins2007considerations}%
  \BibitemOpen
  \bibfield  {author} {\bibinfo {author} {\bibfnamefont {S.~T.}\ \bibnamefont
  {Dawkins}}, \bibinfo {author} {\bibfnamefont {J.~J.}\ \bibnamefont
  {McFerran}}, \ and\ \bibinfo {author} {\bibfnamefont {A.~N.}\ \bibnamefont
  {Luiten}},\ }\href@noop {} {\bibfield  {journal} {\bibinfo  {journal}
  {{{IEEE} Transactions on Ultrasonics, Ferroelectrics, and Frequency
  Control}}\ }\textbf {\bibinfo {volume} {54}},\ \bibinfo {pages} {918}
  (\bibinfo {year} {2007})}\BibitemShut {NoStop}%
\bibitem [{\citenamefont {Benkler}\ \emph {et~al.}(2015)\citenamefont
  {Benkler}, \citenamefont {Lisdat},\ and\ \citenamefont
  {Sterr}}]{benkler2015relation}%
  \BibitemOpen
  \bibfield  {author} {\bibinfo {author} {\bibfnamefont {E.}~\bibnamefont
  {Benkler}}, \bibinfo {author} {\bibfnamefont {C.}~\bibnamefont {Lisdat}}, \
  and\ \bibinfo {author} {\bibfnamefont {U.}~\bibnamefont {Sterr}},\
  }\href@noop {} {\bibfield  {journal} {\bibinfo  {journal} {Metrologia}\
  }\textbf {\bibinfo {volume} {52}},\ \bibinfo {pages} {565} (\bibinfo {year}
  {2015})}\BibitemShut {NoStop}%
\bibitem [{\citenamefont {Rubiola}\ \emph {et~al.}(2005)\citenamefont
  {Rubiola}, \citenamefont {Vernotte},\ and\ \citenamefont
  {Giordano}}]{rubiola}%
  \BibitemOpen
  \bibfield  {author} {\bibinfo {author} {\bibfnamefont {E.}~\bibnamefont
  {Rubiola}}, \bibinfo {author} {\bibfnamefont {F.}~\bibnamefont {Vernotte}}, \
  and\ \bibinfo {author} {\bibfnamefont {V.}~\bibnamefont {Giordano}},\ }in\
  \href {\doibase 10.1109/FREQ.2005.1573901} {\emph {\bibinfo {booktitle}
  {Proceedings of the 2005 IEEE International Frequency Control Symposium and
  Exposition, 2005.}}}\ (\bibinfo {year} {2005})\ pp.\ \bibinfo {pages} {4
  pp.--}\BibitemShut {NoStop}%
\bibitem [{\citenamefont {Best}(2007)}]{best2007phase}%
  \BibitemOpen
  \bibfield  {author} {\bibinfo {author} {\bibfnamefont {R.~E.}\ \bibnamefont
  {Best}},\ }\href@noop {} {\emph {\bibinfo {title} {Phase-locked loops:
  Design, Simulation, and Applications}}}\ (\bibinfo  {publisher} {McGraw-Hill
  Education},\ \bibinfo {year} {2007})\BibitemShut {NoStop}%
\bibitem [{\citenamefont {Demir}(2021)}]{Demir}%
  \BibitemOpen
  \bibfield  {author} {\bibinfo {author} {\bibfnamefont {A.}~\bibnamefont
  {Demir}},\ }\href {\doibase 10.1063/5.0035254} {\bibfield  {journal}
  {\bibinfo  {journal} {Journal of Applied Physics}\ }\textbf {\bibinfo
  {volume} {129}},\ \bibinfo {pages} {044503} (\bibinfo {year}
  {2021})}\BibitemShut {NoStop}%
\bibitem [{\citenamefont {Sadeghi}\ \emph {et~al.}(2020)\citenamefont
  {Sadeghi}, \citenamefont {Demir}, \citenamefont {Villanueva}, \citenamefont
  {K\"ahler},\ and\ \citenamefont {Schmid}}]{Pedram}%
  \BibitemOpen
  \bibfield  {author} {\bibinfo {author} {\bibfnamefont {P.}~\bibnamefont
  {Sadeghi}}, \bibinfo {author} {\bibfnamefont {A.}~\bibnamefont {Demir}},
  \bibinfo {author} {\bibfnamefont {L.~G.}\ \bibnamefont {Villanueva}},
  \bibinfo {author} {\bibfnamefont {H.}~\bibnamefont {K\"ahler}}, \ and\
  \bibinfo {author} {\bibfnamefont {S.}~\bibnamefont {Schmid}},\ }\href
  {\doibase 10.1103/PhysRevB.102.214106} {\bibfield  {journal} {\bibinfo
  {journal} {Phys. Rev. B}\ }\textbf {\bibinfo {volume} {102}},\ \bibinfo
  {pages} {214106} (\bibinfo {year} {2020})}\BibitemShut {NoStop}%
\bibitem [{\citenamefont {Be\ifmmode \check{s}\else
  \v{s}\fi{}i\ifmmode~\acute{c}\else \'{c}\fi{}}\ \emph
  {et~al.}(2023)\citenamefont {Be\ifmmode \check{s}\else
  \v{s}\fi{}i\ifmmode~\acute{c}\else \'{c}\fi{}}, \citenamefont {Demir},
  \citenamefont {Steurer}, \citenamefont {Luhmann},\ and\ \citenamefont
  {Schmid}}]{hajrudinSSO}%
  \BibitemOpen
  \bibfield  {author} {\bibinfo {author} {\bibfnamefont {H.}~\bibnamefont
  {Be\ifmmode \check{s}\else \v{s}\fi{}i\ifmmode~\acute{c}\else \'{c}\fi{}}},
  \bibinfo {author} {\bibfnamefont {A.}~\bibnamefont {Demir}}, \bibinfo
  {author} {\bibfnamefont {J.}~\bibnamefont {Steurer}}, \bibinfo {author}
  {\bibfnamefont {N.}~\bibnamefont {Luhmann}}, \ and\ \bibinfo {author}
  {\bibfnamefont {S.}~\bibnamefont {Schmid}},\ }\href {\doibase
  10.1103/PhysRevApplied.20.024023} {\bibfield  {journal} {\bibinfo  {journal}
  {Phys. Rev. Appl.}\ }\textbf {\bibinfo {volume} {20}},\ \bibinfo {pages}
  {024023} (\bibinfo {year} {2023})}\BibitemShut {NoStop}%
\bibitem [{\citenamefont {Sorden}(1975)}]{hp_paper}%
  \BibitemOpen
  \bibfield  {author} {\bibinfo {author} {\bibfnamefont {J.~L.}\ \bibnamefont
  {Sorden}},\ }\href@noop {} {\bibfield  {journal} {\bibinfo  {journal}
  {Hewlett Packard J}\ }\textbf {\bibinfo {volume} {25}},\ \bibinfo {pages}
  {15} (\bibinfo {year} {1975})}\BibitemShut {NoStop}%
\bibitem [{\citenamefont {Johansson}(2006)}]{johansson2006new}%
  \BibitemOpen
  \bibfield  {author} {\bibinfo {author} {\bibfnamefont {S.}~\bibnamefont
  {Johansson}},\ }in\ \href@noop {} {\emph {\bibinfo {booktitle} {Proceedings
  of the 20th European Frequency and Time Forum}}}\ (\bibinfo {organization}
  {IEEE},\ \bibinfo {year} {2006})\ pp.\ \bibinfo {pages}
  {139--146}\BibitemShut {NoStop}%
\bibitem [{\citenamefont {Lyons}(2010)}]{USigPro}%
  \BibitemOpen
  \bibfield  {author} {\bibinfo {author} {\bibfnamefont {R.~G.}\ \bibnamefont
  {Lyons}},\ }\href@noop {} {\emph {\bibinfo {title} {Understanding Digital
  Signal Processing}}}\ (\bibinfo  {publisher} {Pearson},\ \bibinfo {year}
  {2010})\BibitemShut {NoStop}%
\bibitem [{\citenamefont {Schmid}\ \emph {et~al.}(2023)\citenamefont {Schmid},
  \citenamefont {Villanueva},\ and\ \citenamefont {Roukes}}]{silvanbook2nd}%
  \BibitemOpen
  \bibfield  {author} {\bibinfo {author} {\bibfnamefont {S.}~\bibnamefont
  {Schmid}}, \bibinfo {author} {\bibfnamefont {L.~G.}\ \bibnamefont
  {Villanueva}}, \ and\ \bibinfo {author} {\bibfnamefont {M.~L.}\ \bibnamefont
  {Roukes}},\ }\href@noop {} {\emph {\bibinfo {title} {Fundamentals of
  Nanomechanical Resonators}}},\ \bibinfo {edition} {2nd}\ ed.\ (\bibinfo
  {publisher} {Springer Cham},\ \bibinfo {year} {2023})\BibitemShut {NoStop}%
\bibitem [{\citenamefont {Walls}\ and\ \citenamefont
  {Allan}(1986)}]{Allan_Paper}%
  \BibitemOpen
  \bibfield  {author} {\bibinfo {author} {\bibfnamefont {F.}~\bibnamefont
  {Walls}}\ and\ \bibinfo {author} {\bibfnamefont {D.}~\bibnamefont {Allan}},\
  }\href {\doibase 10.1109/PROC.1986.13429} {\bibfield  {journal} {\bibinfo
  {journal} {Proceedings of the IEEE}\ }\textbf {\bibinfo {volume} {74}},\
  \bibinfo {pages} {162} (\bibinfo {year} {1986})}\BibitemShut {NoStop}%
\end{thebibliography}%

\end{document}